# Enhancement of Tc in BiS$_2$ based superconductors NdO$_{0.7}$F$_{0.3}$BiS$_2$ by substitution of Pb for Bi

S. Demura[1,2], Y. Fujisawa[1], S. Otsuki[1], R. Ishio[1], Y. Takano[2] and H. Sakata[1]

[1]Tokyo University of Science, Department of Physics, Shinjyuku-ku, Tokyo 162-8601, Japan

[2]National Institute for Materials Science, 1-2-1 Sengen, Tsukuba, Ibaraki 305-0047, Japan

E-mail : demura@rs.tus.ac.jp

**Abstract**

We succeed in enhancement of a superconducting transition temperature ($T_c$) for NdO$_{0.7}$F$_{0.3}$BiS$_2$ single crystal by partial substitution of Pb for Bi. The $T_c$ increases with increasing Pb concentration until 6%. The maximum $T_c^{zero}$ is 5.6 K, which is the highest value among BiS$_2$ based superconductors synthesized under an ambient pressure. Pb substitution for Bi induces lattice shrinkage along the $c$ axis. These results reflect that superconductivity in this system is responsive to the lattice strain.

**Introduction**

High temperature superconductivity is the one of the most interesting phenomena in solid state physics. This phenomenon has been achieved in cuprates and iron-based superconductors [1-13]. One of the common features in these materials is the layered structure composed of block layers and conduction layers. This structure leads to two dimensional conduction, which is intrinsic to unconventional superconducting mechanism. This structure also contribute to the carrier doping which is needed to realize superconductivity: the block layers supply carriers into the conduction layers without introducing structural disorder of the conduction layers.

Recently, $BiS_2$ based superconductors has been discovered [14,15]. These superconductors have the similar layered structure to cuprates and iron-based superconductors. So, this material is possible candidate for another high temperature superconductor. Up to now, many members of $BiS_2$ superconductors have been reported [16-26]. Among them, $NdOBiS_2$ shows superconductivity when a part of $O^{2-}$ ions are partly substituted $F^-$ ions [16]. The $T_c$ changes by the control of the F concentration and becomes the maximum value of 5.2K at the F concentration of 40 %. This $T_c$ is the highest value among $BiS_2$ based superconductors synthesized in an ambient pressure. Interestingly, the $T_c$ increases up to around 6.5 K when the samples are measured under a high pressure [27]. This increase of $T_c$ is common properties in $BiS_2$ compounds with the block layers of $Ln$O ($Ln$=La, Pr, Ce, Nd) [28-36].

Furthermore, a theoretical study for La(O,F)BiS$_2$ using a first principle band calculation indicated that the distance between La and S outside of the BiS$_2$ layer significantly affects the electronic structure [37]. These results may imply that the properties of BiS$_2$ superconductors are quite sensitive for a lattice strain. Considering these facts, it is interesting to induce the lattice strain into the compounds by a partial substitution of atoms to investigate the relation between $T_c$ and the lattice strain.

In this study, we report an increase of $T_c$ for NdO$_{0.7}$F$_{0.3}$BiS$_2$ single crystals by partial substitution of Pb for Bi. The ionic radius of Pb ion is larger than that of Bi ion. Therefore, a strain can be introduced to the crystal structure in an ambient pressure when Pb ions are substituted for Bi ions. These single crystals were synthesized by a flux method in quarts tubes. A $c$ axis significantly decreases with increasing Pb concentration up to 6% while a $a$ axis gradually increases. The $T_c$ of these crystals increases with increasing Pb concentration until 6 %. A maximum value of $T_c^{zero}$ is 5.6 K, which is the highest value among those of BiS$_2$ superconductors synthesized in an ambient pressure. These facts suggest that superconducting properties are responsible to the lattice strain.

**Experimental**

Single crystal samples of NdO$_{0.7}$F$_{0.3}$Bi$_{1-x}$Pb$_x$S$_2$ ($x$=0.01-0.1) were prepared by a CsCl flux method in vacuumed quarts tubes [38,39]. Mixtures of Bi (Mituswa Chemicals Co. Ltd., 99.9%), $Bi_2S_3$, $Bi_2O_3$ (Kojyundo Chemical Laboratory Co. Ltd., 99.99%), $BiF_3$ (Stella Chemifa Co. Ltd., 99%), $Nd_2S_3$ (Kojyundo Chemical Laboratory Co. Ltd., 99%), and PbO (Kojyundo Chemical Laboratory Co. Ltd., 99.9%) were ground with nominal compositions of $NdO_{0.7}F_{0.3}Bi_{1-x}Pb_xS_2$ ($x$=0.01-0.1). $Bi_2S_3$ was obtained by sintering the mixtures of Bi and S (Kojyundo Chemical Laboratory Co. Ltd., 99.99%) in the evacuated quartz tube at 500 °C for 10 hours. The mixture of 0.8 g was mixed with CsCl powder (Kojyundo Chemical Laboratory Co. Ltd., 99.9%) of 5 g, and sealed in an evacuated quartz tube. The tube was heated at 800 °C for 10 hours and cooled down to 630 °C at a rate of 0.5 or 1.0 °C/h. After this thermal process, the sintered materials were washed by distilled water to remove the flux. The obtained single crystals were characterized by X-ray diffraction measurements with Cu-K$\alpha$ radiation using the $\theta$-$2\theta$ method. The surface structure of single crystals was observed by a laboratory-build scanning tunneling microscope (STM) at 4 K. The temperature dependence of magnetic susceptibility was measured by a superconducting quantum interface device (SQUID) magnetometer with an applied field of 10 Oe. The field was applied parallel to the $c$ axis of the sample. $T_c^{mag}$ is defined as the crossing point of fitting lines for the magnetic susceptibility in the normal state near the transition and the line in the drop area during the transition. The resistivity measurements were performed using the four-terminal method from 2 to 300 K. The $T_c^{onset}$ is defined as a crossing point of fitting lines for resistivity in the normal state near the transition and the line in the drop area during the transition. The $T_c^{zero}$ is defined as the temperature where resistivity becomes zero.

**Results**

Figure 1(a) shows X-ray diffraction profiles for the single crystal samples of NdO$_{0.7}$F$_{0.3}$Bi$_{1-x}$Pb$_x$S$_2$ ($x$=0.01-0.10), respectively. The bottom profile shows (00$l$) peaks of NdOBiS$_2$ as reference. All of the peaks are corresponded to the (00$l$) peaks of CeOBiS$_2$ type structure with the space group *P*4/*nmm* symmetry. Enlarged (004) peaks are shown in Fig.1(b). These peaks gradually shift to higher angle with increasing Pb concentration. Figure 2(a) exhibits $c$ axis of Pb doped samples. The $c$ axis significantly decreases with increasing Pb concentration until $x$=0.06, and become almost constant up to $x$=0.10. The $a$ axis was estimated by X-ray diffraction using the powder prepared by grinding the single crystals. As shown in Fig. 2(b), the $a$ axis gradually increases with increasing Pb concentration up to $x$=0.10. The monotonic change in the $a$ axis guarantees the partial substitution of Pb for Bi in this $x$ range.

We also observed the cleaved sample surface by STM technique. Figure 3 shows a STM image of NdO$_{0.7}$F$_{0.3}$Bi$_{0.96}$Pb$_{0.04}$S$_2$ single crystal. The observed surface shows a square lattice with a lattice constant of 3.9 Å, which corresponds to the lattice constant $a$. No indication of Pb segregation or phase separation was observed. The observed atoms are thought to be Bi atoms on the topmost BiS$_2$ layer as in the case of NdO$_{0.7}$F$_{0.3}$BiS$_2$ [40]. The number of dark spots, which indicate Bi defect, was about 4%. This concentration of the defect is the almost same as previous report in NdO$_{0.7}$F$_{0.3}$BiS$_2$ [40]. Thus, the number of defects is not affected by the Pb substitution.

Figure 4(a) exhibits temperature dependences of magnetic susceptibility for NdO$_{0.7}$F$_{0.3}$Bi$_{1-x}$Pb$_x$S$_2$ ($x$=0.01-0.10) single crystals. Large diamagnetic signal corresponds to superconductivity can be seen for all Pb doped samples. To determine these superconducting transition temperatures ($T_c$), the magnetic susceptibility near the $T_c$ is shown in Fig. 4(b). The $T_c$ increases with increasing Pb concentration until $x$=0.06, and decreases at $x$=0.10. The highest $T_c^{mag}$ is 5.6 K.

The temperature dependence of resistivity for all Pb doped samples is presented in Fig. 5(a). All samples show the sharp superconducting transition and zero resistivity around 6K, which is correspond to superconducting transition. The resistivity near the $T_c$ is exhibited in Fig. 5(b). The $T_c^{onset}$ and $T_c^{zero}$ increase with increasing Pb concentration until $x$=0.06, and decrease at $x$=0.10. The maximum $T_c^{zero}$ is 5.6 K at $x$=0.06, which is the highest value among the BiS$_2$-based superconductors synthesized under an ambient pressure.

In order to discuss the temperature dependence of resistivity, we fit the resistivity data to the formula:

$$\rho=\rho_0+AT^n$$

where $\rho_0$ and $A$, n is fitting parameters. These fitting lines are drown as the gray lines in the Fig. 5(a) and fitting parameters are shown in a table 1. The resistivity of samples with $x$=0.02 and 0.04 can be fitted to the formula from 300 K down to 80 K, and shows a slight upturn below 80 K. For the samples with $x$=0.06 and 0.10, the fitting region goes down to 20 K. Namely, these sample shows the more metallic behavior than the samples with $x$=0.02 and 0.04. This result implies that the upturn of resistivity is suppressed by Pb doping.

The resistivity at the room temperature for each Pb doped sample increases with increasing Pb concentration until $x$=0.06. This change in resistivity mainly comes from the change in $\rho_0$. Because Pb substitution of Bi in $BiS_2$ layer, which is the conduction layer, may increases the disorder of the conduction plane, residual resistivity $\rho_0$ seems to increase with Pb doping. However, parameter $A$ also increases at $x$=0.06. The increase in $A$ cannot be explained by the disorder. This indicates that the Pb doping induces not only as enhancement of the disorder but also modification of the electronic structure through the change in the lattice parameter. It is noted that the resistivity at the room temperature slightly decreases when the amount of Pb is $x$=0.10. This also shows that the effect of the Pb doping is not only the enhancement of the disorder.

Figure 6 shows a Pb concentration dependence of $T_c$ obtained by each measurement. All $T_c$ gradually increase with increasing Pb concentration until $x$=0.06. This result indicates that superconductivity is obviously enhanced by the Pb doping. Now, let's consider a relation between Pb doping and the enhancement of $T_c$. Following two effects are expected by Pb doping. One is the change of the carrier number. A valence of Pb ion is typically $Pb^{2+}$. This valence is one electron fewer than that of $Bi^{3+}$. This means that carriers are taken from the compound by Pb doping. Thus, the enhancement of $T_c$ may due to the change in the carrier number. However, this is not the case. In the previous study, the carrier number dependence on $T_c$ was reported in $NdO_{1-x}F_xBiS_2$, where the carrier number was controlled by the amount of $F^-$ concentration substituted for $O^{2-}$. In that study, the maximum $T_c$ was achieved at the F concentration of $x=0.4$ [16]. Thus, to obtain the maximum $T_c$ by carrier doping into $NdO_{0.7}F_{0.3}BiS_2$, it needs 0.1 electron per Bi ion doping into the sample. On the other hand, Pb doping introduces the hole carriers into the sample. This should decrease $T_c$. Therefore, the enhancement of $T_c$ is not due to the carrier doping. Furthermore, the observed maximum $T_c$ in Pb doped sample is higher than that of $NdO_{1-x}F_xBiS_2$. Thus, the observed enhancement of $T_c$ cannot be explained by the change in the carrier number.

The other effect of the Pb doping is the change in the lattice constant. Ionic radius of $Pb^{2+}$ ion is larger than that of $Bi^{3+}$ ion. Thus, the crystal is slightly strained by doping Pb ions. As shown in Fig. 2, the lattice constant $a$ and $c$ changes depending on the amount of doped Pb. This effect may be related to the enhancement of $T_c$. Actually, many papers have revealed that $T_c$ of $BiS_2$ based superconductors with $Ln$O$BiS_2$ type structure increases under the high-pressure where the lattice constant decreases [27-36]. Theoretical study also reported that the band structure of BiS$_2$-based materials is sensitive to the lattice constant along the $c$ axis [37,41,42]. Thus, it is plausible that the $T_c$ enhancement with Pb doping is achieved by shrinkage along the $c$ axis. In this scenario, the saturation of $T_c$ at $x$=0.1 is consistent with the saturation of the lattice constant $c$ as shown in Fig.2 (b).

**Summary**

We achieved the increase of $T_c$ by the Pb substitution of Bi in NdO$_{0.7}$F$_{0.3}$BiS$_2$. The enhancement of $T_c$ is attributed the change in the lattice contraction along the $c$ axis. The $T_c^{zero}$ increases with decreasing $c$ axis, and becomes the maximum value around 5.6 K at the Pb concentration of 6 %, which is the highest value among BiS$_2$ based superconductors synthesized under the ambient pressure. These results indicate that superconductivity in this material is strongly associated with the lattice strain.

**Acknowledgements**

This work was partly supported by a Grant-in-Aid for Scientific Research from the Ministry of Education, Culture, Sports, Science and Technology (KAKENHI).

**Figure caption**

Fig. 1 (Color online)

(a) X-ray diffraction patterns for single crystals of $NdO_{0.7}F_{0.3}Bi_{1-x}Pb_xS_2$ ($x$=0.01-0.1). The bottom black line shows the reference (00*l*) peaks of $NdOBiS_2$. (b) Enlarged figure of (004) peaks for Pb doped samples.

Fig. 2 (Color online)

(a, b) Pb concentration dependence of *c* and *a* axis. These *a* axis were estimated by the powder X-ray measurements using ground single crystals.

Fig. 3(Color online)

STM image of the $NdO_{0.7}F_{0.3}Bi_{0.96}Pb_{0.04}S_2$ single crystal on a $130 \times 130$ Å$^2$ field of view taken at $V = +1400$ mV and $I = 500$ pA. The inset show a Fourier transform image. The red arrows indicate the direction an atom period corresponded to cell parameter of *a* axis.

Fig. 4 (Color online)

(a) The temperature dependence of the magnetic susceptibility for single crystals of $NdO_{0.7}F_{0.3}Bi_{1-x}Pb_xS_2$ ($x$=0.01-0.1). (b) The enlarged figure of magnetic susceptibility near the superconducting transition.

Fig. 5 (Color online)

(a) The temperature dependence of resistivity for single crystals of $NdO_{0.7}F_{0.3}Bi_{1-x}Pb_xS_2$ ($x$=0.01-0.1) between 300 and 2 K. (b) The enlarged picture of resistivity between 10 and 2 K.

Fig. 6 (Color online)

The nominal F concentration $x$ - superconducting transition temperature phase diagram for single crystals of $NdO_{0.7}F_{0.3}Bi_{1-x}Pb_xS_2$ ($x$=0.01-0.1).

Table 1

Fitting parameters of resistivity for single crystals of $NdO_{0.7}F_{0.3}Bi_{1-x}Pb_xS_2$ ($x$=0.02-0.1).

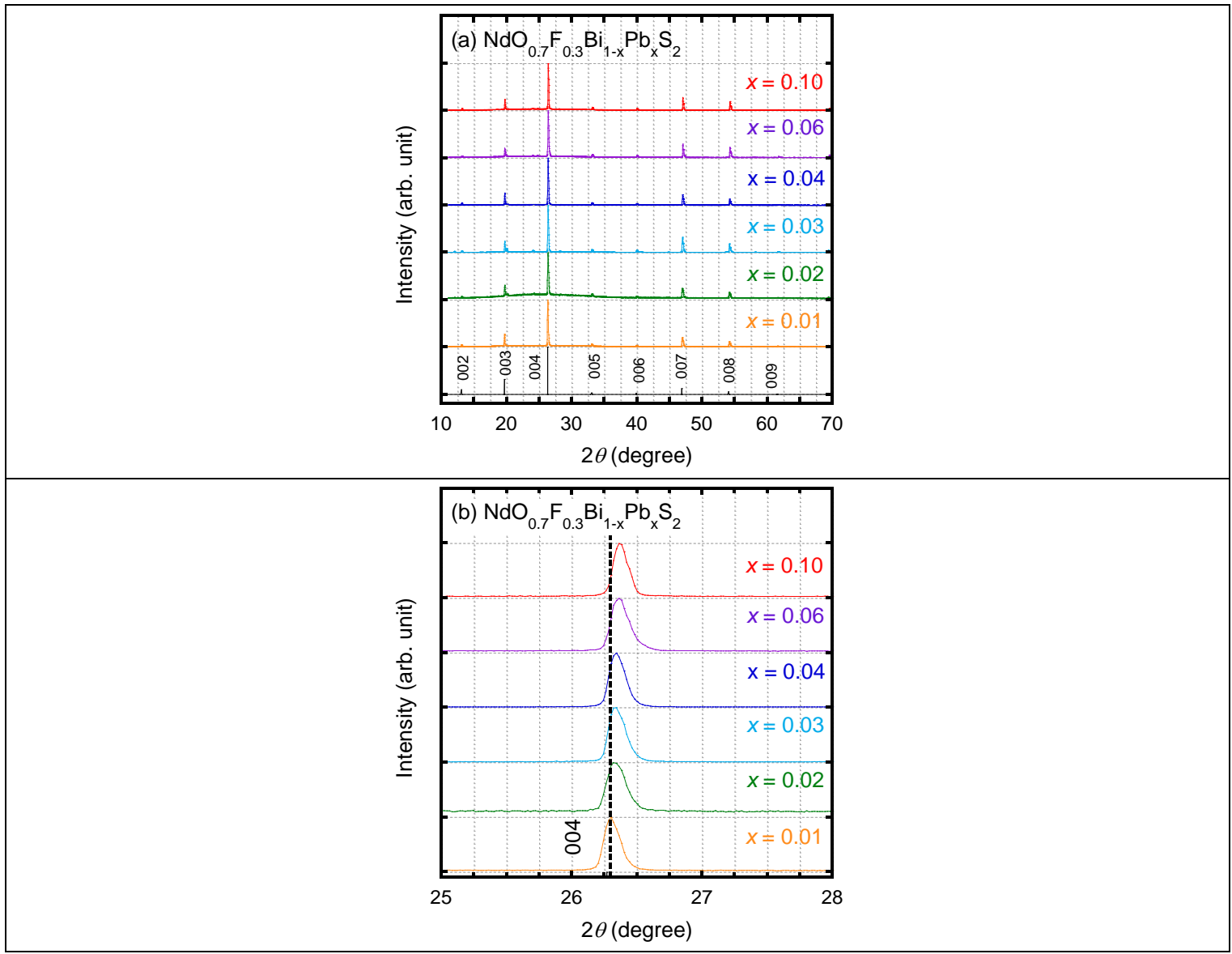

Fig. 1(a, b)

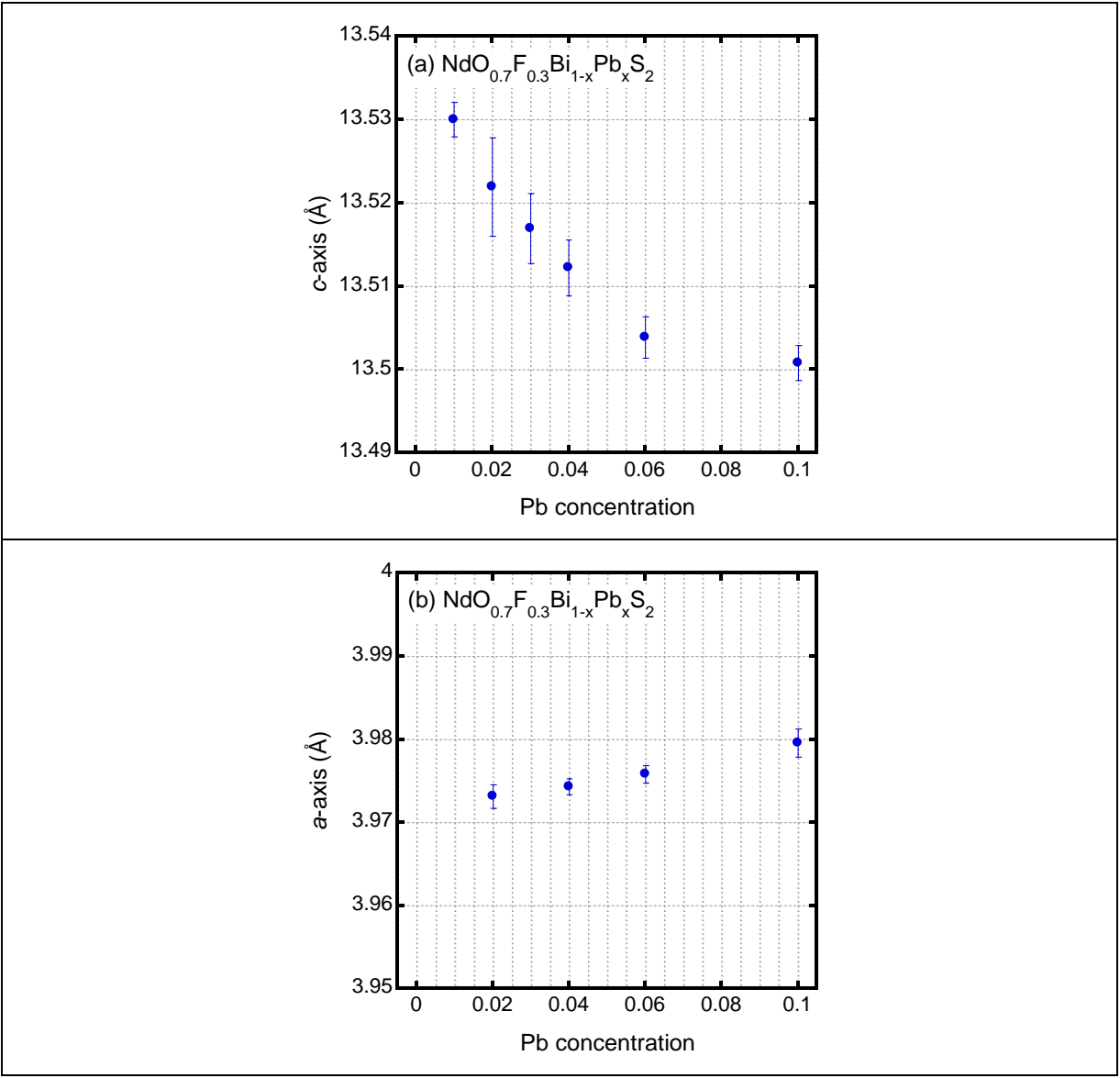

Fig. 2(a,b)

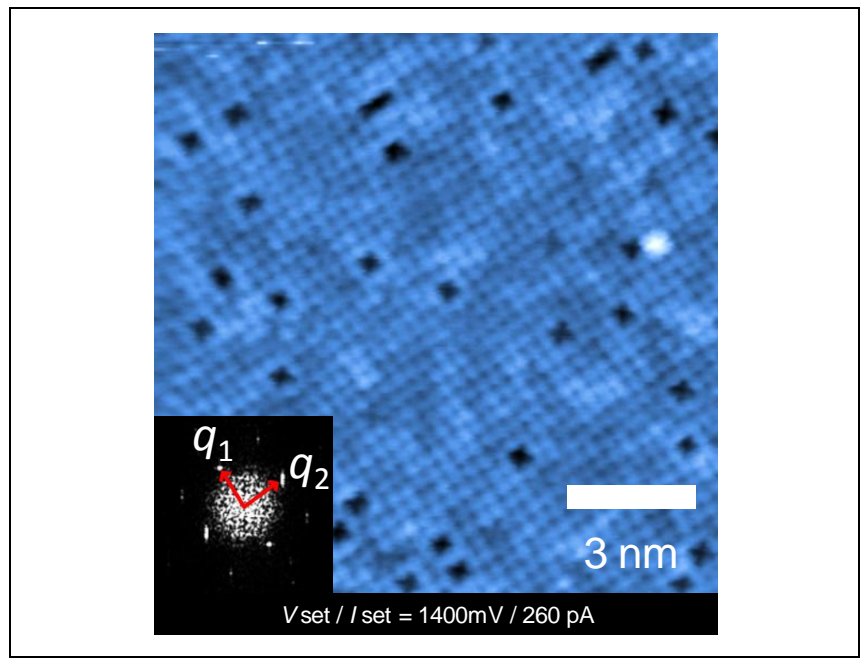

Fig. 3

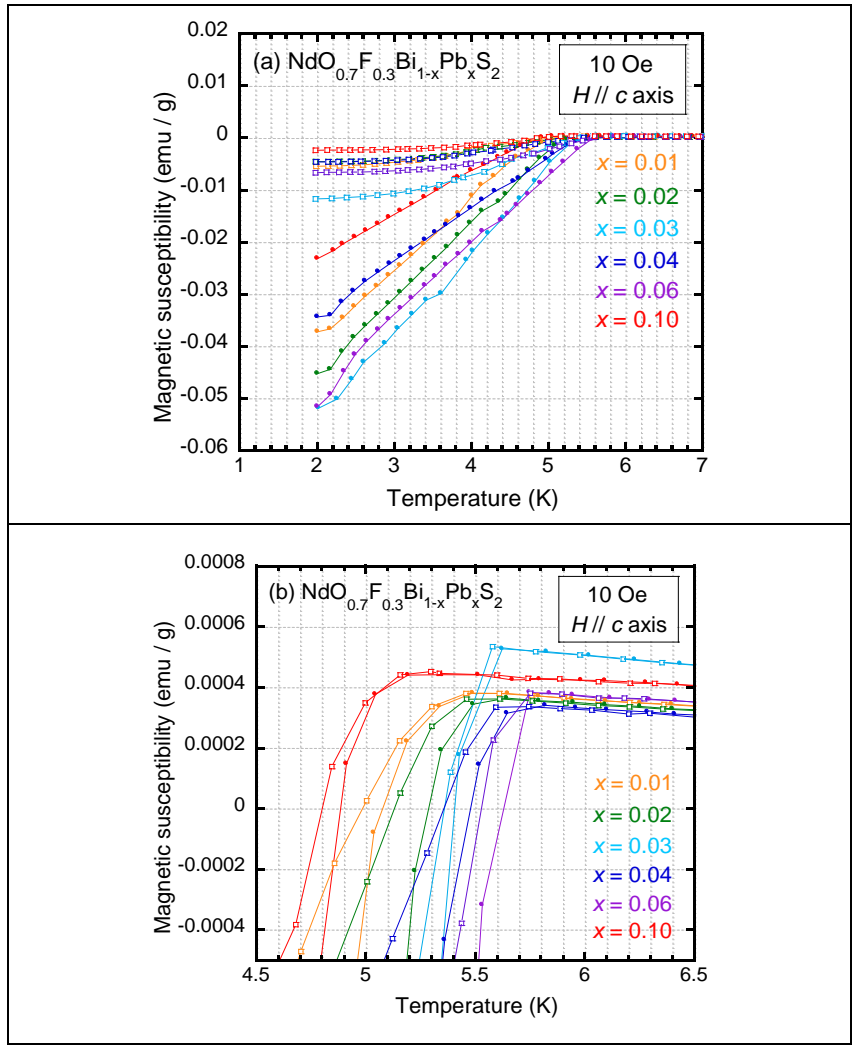

Fig. 4(a,b)

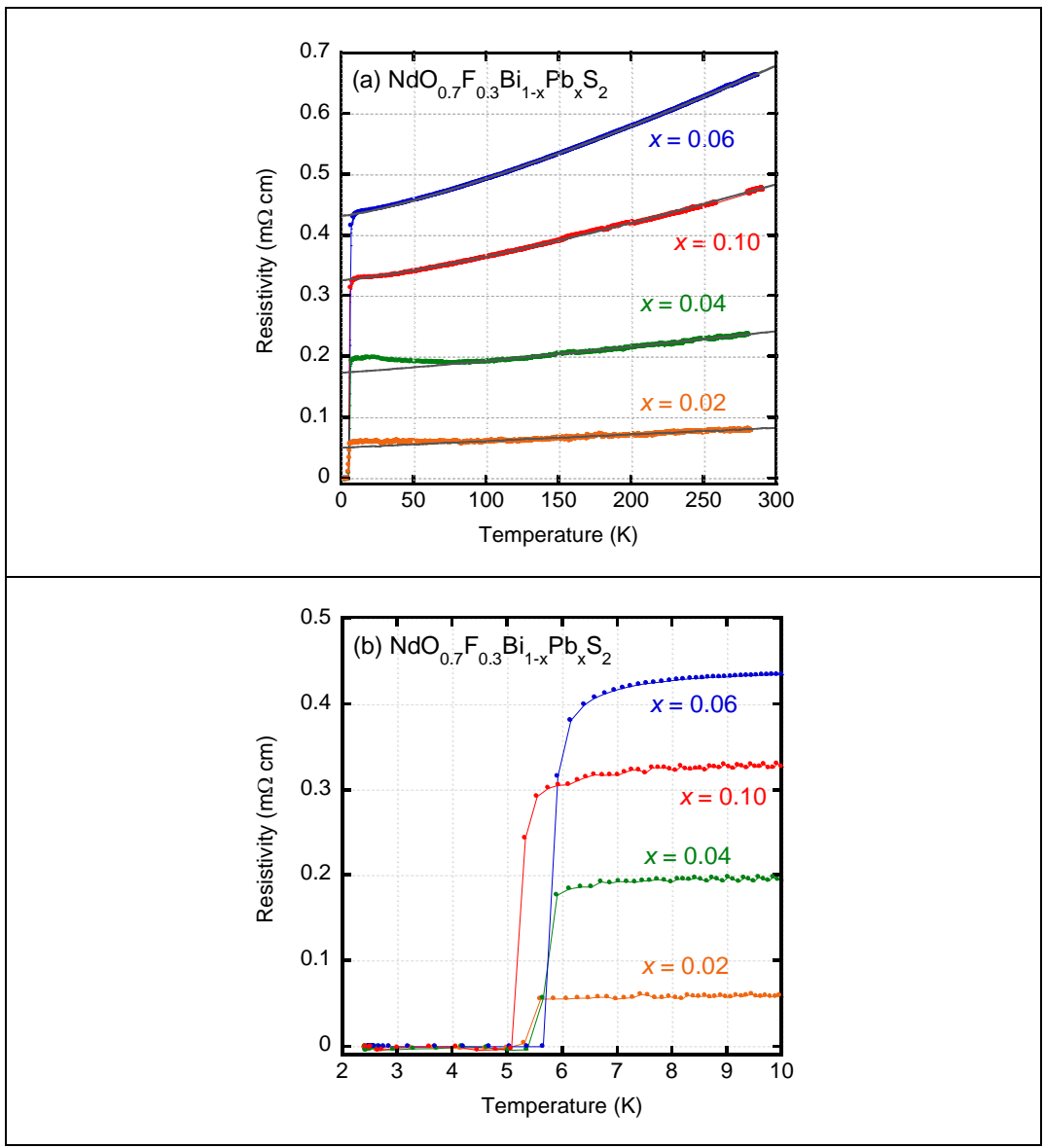

Fig. 5(a, b)

| F concentration | 0.02 | 0.04 | 0.06 | 0.1 |
|---|---|---|---|---|
| $\rho_o$ ($\times 10^{-1}$ Ω cm) | 0.49 | 1.73 | 4.31 | 3.25 |
| $A$ ($\times 10^{-4}$) | 1.20 | 0.94 | 1.76 | 1.08 |
| $n$ | 0.98 | 1.15 | 1.27 | 1.28 |

Table 1

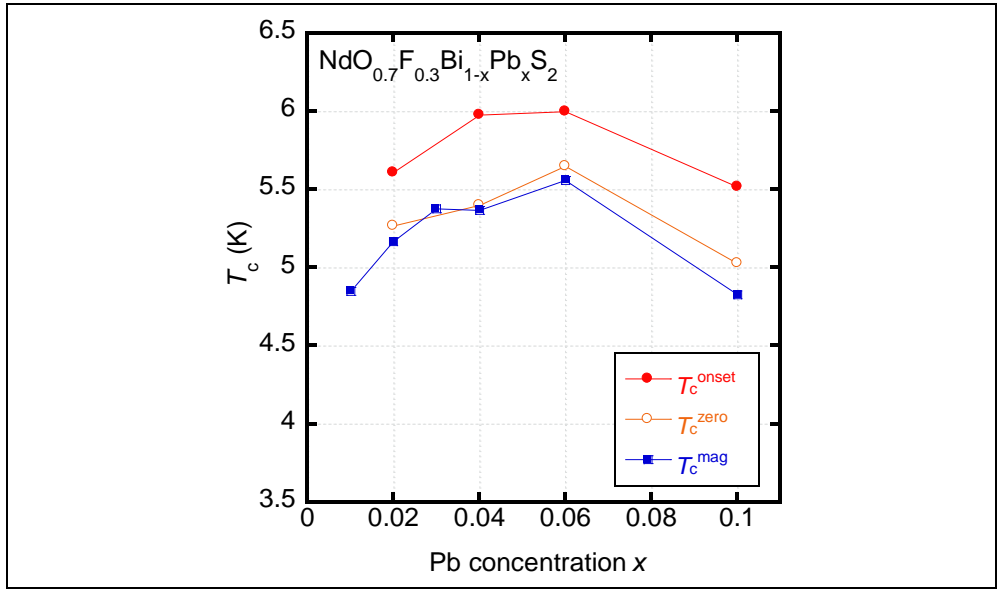

Fig. 6